\RequirePackage[T1]{fontenc}
\documentclass[journal,twoside,web]{ieeecolor}
\usepackage{generic}
\def\logoname{nologo}
\definecolor{subsectioncolor}{rgb}{0,0.541,0.855}
\makeatletter
\def\subsection{\@startsection{subsection}{2}{\z@}{3.5ex plus 1.5ex minus 1.5ex}%
{0.7ex plus .5ex minus 0ex}{\everypar{\color{subsectioncolor}}\normalsize\normalfont\everymath={\sf}\sf\itshape\raggedright}}%
\makeatother
\usepackage{amsmath,amssymb,amsfonts}
\usepackage{graphicx}
\usepackage{booktabs,multirow}
\usepackage{url}
\usepackage{etoolbox}
\usepackage{algorithm,algorithmic}
\newcommand{\vx}{\boldsymbol{x}}\newcommand{\vy}{\boldsymbol{y}}\newcommand{\vk}{\boldsymbol{k}}\newcommand{\vg}{\boldsymbol{g}}\newcommand{\ve}{\boldsymbol{e}}\newcommand{\vL}{\boldsymbol{L}}
\newcommand{\mA}{\boldsymbol{A}}\newcommand{\mM}{\boldsymbol{M}}\newcommand{\mF}{\boldsymbol{F}}\newcommand{\mS}{\boldsymbol{S}}

\makeatletter\patchcmd{\thebibliography}{plus pt}{plus 0pt}{}{\errmessage{thebibliography patch failed}}\makeatother
\makeatletter\let\NAT@parse\undefined\makeatother
\usepackage[hidelinks]{hyperref}
\begin{document}
\title{One Block Is Enough: LoopMR for Accelerated MRI Reconstruction}
\author{Yimeng~He, Yibin~Xie, Yi~Zheng, Yifan~Gao, Lixia~Wang, Yin-Chen~Hsu, Yufeng~Wang, Meng~Lu,\\ Debiao~Li, and Xiuzhen~Huang
\thanks{(Corresponding authors: Debiao Li and Xiuzhen Huang.)}
\thanks{All authors are with Cedars-Sinai Medical Center, Los Angeles, CA 90048 USA. Yimeng He is with the Biomedical Imaging
Research Institute and the Department of Computational Biomedicine, Xiuzhen Huang is with the Department of Computational
Biomedicine, and the other authors are with the Biomedical Imaging Research Institute (e-mail: Yimeng.He@cshs.org,
Debiao.Li@csmc.edu, Xiuzhen.Huang@csmc.edu).}}
\maketitle

\begin{abstract}
Deep-learning unrolled networks are the leading approach to accelerated MRI reconstruction. State-of-the-art models such as
PromptMR+ learn a distinct block for every cascade, so their size grows with the number of cascades. Sharing one block across iterations avoids adding
parameters, but earlier shared-weight models typically used a tiny block and produced worse images than models with distinct weights.
Therefore, it remained unclear whether weight sharing itself or the choice of block was the problem. Recent advances in looped
transformers, which reuse the same block across layers, show that they can match or outperform their distinct-weight counterparts on several tasks. Motivated by this, we propose LoopMR, which reuses a single Restormer block at every iteration as its core. Besides
the current image estimate, the block receives data-consistency inputs and latent images. On four datasets covering the brain
(CC359), knee (fastMRI), heart (CMRxRecon2023) and carotid (in-house), LoopMR significantly outperforms E2E-VarNet, PromptMR and PromptMR+ in PSNR and SSIM at every acceleration
factor ($p < 0.01$), with about 1/9 of PromptMR+'s parameters (10.3M versus 92.9M). It also keeps its lead at unseen
acceleration factors and on an unseen dataset. With our transformer block, the shared model matches or outperforms its
distinct-weight counterpart at 6, 12 and 24 iterations. At a fixed number of iterations, shared transformer blocks match or outperform their distinct-weight versions, while shared CNN blocks
perform worse. These results suggest that looping one
well-chosen block can outperform stacking distinct cascades in MRI reconstruction, with a fraction of the parameters.
\end{abstract}


\section{Introduction}
MRI is one of the most widely used tools in clinical diagnosis and research. It provides excellent soft-tissue contrast without
radiation. However, due to physical constraints, acquiring a fully sampled scan is slow, which increases costs for patients and makes
the images prone to motion artifacts. Consequently, reducing scan time is a valuable research area with a profound clinical impact.
Traditional acceleration methods include parallel imaging (PI; e.g., SENSE~\cite{pruessmann1999sense}, GRAPPA~\cite{griswold2002grappa})
and compressed sensing (CS)~\cite{lustig2007sparse}, but the acceleration they achieve is modest, and some of them require specific
acquisition protocols.

Deep learning reconstruction is a popular topic in MRI acceleration research~\cite{knoll2020survey} and already has real-world
applications such as Siemens Deep Resolve~\cite{siemens_deepresolve}. Since classical reconstruction algorithms are based on iterative
improvement, the unrolled network received the most attention and is proven to be one of the most effective ones~\cite{monga2021unrolling}.
Early work such as MoDL~\cite{aggarwal2019modl} used a single network with shared weights for the repeated refinement. Later
shared-weight models include deep equilibrium networks (DnCNN)~\cite{gilton2021deq}, which iterate one network to a fixed point, and
ReconFormer~\cite{guo2024reconformer}, which reuses one transformer across iterations. However, these shared-weight models are
typically small: MoDL has about 0.1M parameters, DnCNN has 0.56M,
and ReconFormer has 1.1M in total. Their
performance was increasingly surpassed by cascaded models with distinct weights. Deep Cascade~\cite{schlemper2018cascade} and
VarNet~\cite{hammernik2018varnet} first used this distinct design, and later models such as E2E-VarNet~\cite{sriram2020e2e},
PromptMR~\cite{xin2024promptmr} and PromptMR+~\cite{xin2024promptmrplus} further improved accuracy by stacking more of these
cascades. PromptMR also won first place in both tasks of the CMRxRecon2023 challenge~\cite{lyu2025cmrxrecon2023}, and PromptMR+ won first place in both tasks of the CMRxRecon2024 challenge~\cite{cmrxrecon2024_leaderboard}. Distinct weights have thus
become the de facto choice for MRI reconstruction. However, this design has a cost: the number of parameters grows
with every cascade. PromptMR+ already has 92.9M parameters, and the more recent SDUM~\cite{wang2025sdum}, which reports gains over
PromptMR+, stacks 18 Restormer-based cascades with distinct weights, reaching about 759M.

Recent work on large language models (LLMs) suggests that the number of parameters does not have to grow with depth.
Looped transformers~\cite{dehghani2019universal}, which reuse a single block across depth, have shown strong promise. A $k$-layer
transformer looped $L$ times nearly matches a $kL$-layer model on reasoning tasks~\cite{saunshi2025reasoning}, and pre-trained LLMs
converted to a single looped block outperform vanilla models of equal size~\cite{bae2025relaxed}. It also shows stability and follows
predictable scaling laws beyond parameters and data~\cite{prairie2026parcae}. These gains concentrate on problems that require
iterative refining rather than storing knowledge~\cite{saunshi2025reasoning}, which suits MRI reconstruction well. Classical iterative
algorithms, such as gradient descent and ADMM~\cite{boyd2011admm}, apply the same update with the same image prior at every iteration.
Only the estimate changes, and data consistency (DC) brings the measured k-space back in at each step. A shared block keeps this
structure: what changes between iterations is carried by its inputs rather than by its weights. Since earlier shared-weight models
were small, whether a single looped transformer can achieve comparable or better performance than the state-of-the-art
distinct-weight model is still a question to be tested.

\begin{figure*}[!t]
\centering
\includegraphics[width=\textwidth]{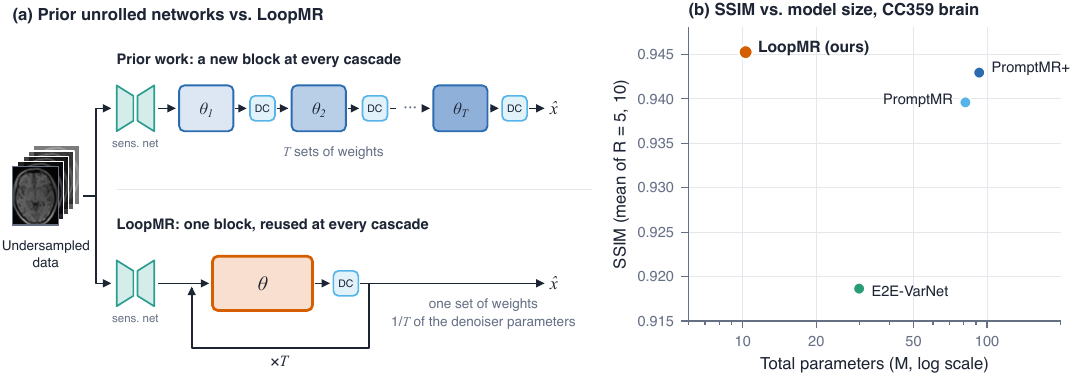}
\caption{(a) Prior unrolled networks learn a new block at every cascade; LoopMR reuses one block at every iteration,
with $1/T$ of the denoiser parameters. (b) SSIM (mean of $R = 5, 10$) against total parameters on CC359 brain.}
\label{fig:teaser}
\end{figure*}

In this work, we introduce LoopMR, an unrolled reconstruction network that reuses a single transformer block at every iteration
instead of learning a distinct block for each iteration (Fig.~\ref{fig:teaser}a), and we test this question with controlled shared-versus-distinct
comparisons. We deliberately keep the design simple: the shared block is a Restormer~\cite{zamir2022restormer} with four additional
PromptMR+-style inputs (i.e., three DC inputs and a set of latent states), so that most of the gain comes from the looping design
rather than from new modules. Our contributions are:
\begin{itemize}
\item \textbf{LoopMR: state-of-the-art accuracy with one block.} To our knowledge, LoopMR is the first shared-weight unrolled
network to outperform a state-of-the-art distinct-weight model. It loops one Restormer block with data-consistency inputs
and latents. It significantly outperforms PromptMR+, PromptMR and E2E-VarNet on brain, knee, heart and carotid data, with about 1/9 of PromptMR+'s parameters (Fig.~\ref{fig:teaser}b). It keeps its lead at unseen acceleration factors
and on an unseen dataset.
\item \textbf{A shared block matches distinct blocks at every depth.} A shared transformer block matches or outperforms an identical
model with distinct weights at every depth we tested, with far fewer parameters.
\item \textbf{Sharing depends on the block.} Shared transformer blocks match or outperform their distinct-weight counterparts,
whereas shared CNN blocks perform worse. A larger block widens the transformer's advantage and narrows the CNN's gap. This may
explain why earlier shared-weight models, which used tiny CNN blocks, performed worse than distinct-weight models.
\end{itemize}

\section{Method}
\subsection{Problem Formulation}
We can estimate the complex image $\vx$ from its undersampled k-space measurements $\vy$ by solving
\begin{equation}
\min_{\vx} \ \tfrac{1}{2} \| \mA \vx - \vy \|_2^2 + \mathcal{R}(\vx), \label{eq:obj}
\end{equation}
where $\mA = \mM \mF \mS$ is the forward process of acquiring the k-space measurements: $\mS$ maps the image to coil images with the
sensitivity maps, $\mF$ is the Fourier transform and $\mM$ is the
undersampling mask. The first term is the data-consistency term, which keeps the estimate faithful to the measurements, and the
regularizer $\mathcal{R}$ promotes image priors such as sparsity~\cite{lustig2007sparse}. $\mA^H \vy$ is the zero-filled image, and
$\mA^H \mA \vx$ re-undersamples an image with the acquired mask. Eq.~(\ref{eq:obj}) can be solved iteratively by gradient descent,
\begin{equation}
\vx_{t+1} = \vx_t - \eta_t\, \vg_t, \qquad \vg_t = \mA^H (\mA \vx_t - \vy) + \nabla \mathcal{R}(\vx_t), \label{eq:gd}
\end{equation}
where $\nabla\mathcal{R}(\vx_t)$ is the gradient or the proximal mapping of $\mathcal{R}$~\cite{boyd2011admm}, $\eta_t$ is the step
size and $t \in \{0, 1, \dots, T-1\}$.  Deep unrolled models truncate (\ref{eq:gd}) to $T$ iterations and replace $\nabla\mathcal{R}(\vx_t)$ with a network learned from pairs of undersampled and fully
sampled data. Cascaded models such as VarNet~\cite{hammernik2018varnet}, E2E-VarNet~\cite{sriram2020e2e} and
PromptMR+~\cite{xin2024promptmrplus} learn a distinct network for every iteration. Shared-weight models such as
MoDL~\cite{aggarwal2019modl} use one network at every iteration, as (\ref{eq:gd}) does.

\subsection{The Loop}
\begin{figure*}[!t]
\centering
\includegraphics[width=\textwidth]{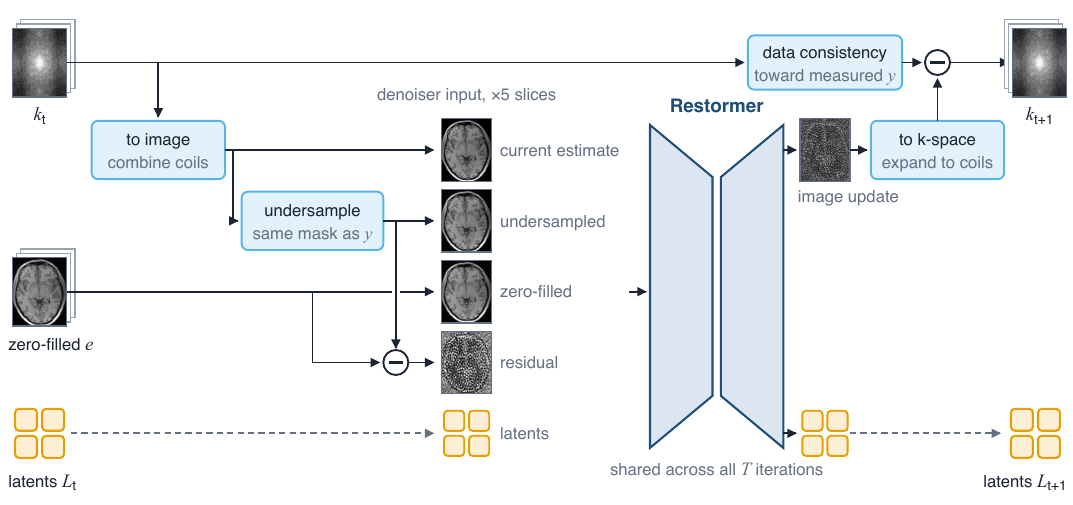}
\caption{One LoopMR iteration. The shared Restormer receives the current estimate, its undersampled version, the
zero-filled image, their residual and the latents, and returns an image update and new latents. Data consistency
pulls the k-space toward the measurements. The same block is used at all $T$ iterations.}
\label{fig:method}
\end{figure*}

Algorithm~\ref{alg:loopmr} gives one forward pass of LoopMR. The block takes two sets of inputs: the current image estimate and
a set of latent images. Both start from the zero-filled reconstruction: the state is initialized with the measured k-space,
$\vk_0 = \vy$, so the first estimate $\vx_0$ is the zero-filled image, and the latents $\vL_0$ are $K$ copies of it
(line~3). At each iteration, we coil-combine the k-space state into images (line~5) and compute the data-consistency terms: the
re-undersampled estimate $\vg_t = \mA^H \mA \vx_t$, the zero-filled image $\ve$ and their residual $\vg_t - \ve$ (line~6). These are
passed to the block together with the images and the latents (line~7). The block works in the image domain and returns an image
update and new latents. At the end of the iteration, the update is converted back to k-space and combined with a soft
data-consistency step (line~8), as in E2E-VarNet~\cite{sriram2020e2e}; this is the gradient step (\ref{eq:gd}) taken in k-space,
with the block's update in place of $\nabla\mathcal{R}(\vx_t)$. The same block and the same learned scalars $\alpha$ and $\eta$
are used at every iteration. Fig.~\ref{fig:method} shows one iteration.

\begin{algorithm}[!t]
\caption{LoopMR reconstruction}\label{alg:loopmr}
\begin{algorithmic}[1]
\REQUIRE undersampled k-space $\vy$, mask $\mM$, number of iterations $T$, number of latents $K$
\ENSURE reconstruction $\hat{\vx}$
\STATE $\mS \leftarrow \mathrm{SME}(\vy)$ \COMMENT{maps from the calibration region}
\STATE $\ve \leftarrow \mA^H \vy$ \COMMENT{zero-filled image}
\STATE $\vk_0 \leftarrow \vy$; \ $\vL_0 \leftarrow$ $K$ copies of $\ve$
\FOR{$t = 0$ \TO $T-1$}
\STATE $\vx_t \leftarrow \mS^H \mF^{-1} \vk_t$ \COMMENT{coil-combine}
\STATE $\vg_t \leftarrow \mA^H \mA\, \vx_t$ \COMMENT{re-undersample}
\STATE $(\Delta \vx_t, \vL_{t+1}) \leftarrow D_\theta(\vx_t, \vg_t, \vL_t, \ve, \vg_t - \ve)$
\STATE $\vk_{t+1} \leftarrow \vk_t - \eta\, \mM(\vk_t - \vy) - \alpha\, \mF \mS\, \Delta \vx_t$
\ENDFOR
\RETURN $\hat{\vx} = \mathrm{RSS}(\mF^{-1} \vk_T)$ \COMMENT{root sum of squares~\cite{roemer1990phased}}
\end{algorithmic}
\end{algorithm}

\subsection{The Shared Block}
Inspired by SDUM~\cite{wang2025sdum}, which uses a Restormer~\cite{zamir2022restormer} as the denoiser of each cascade, we use a
Restormer as the shared block $D_\theta$. Restormer computes self-attention across channels rather than across pixels, so its cost
grows linearly with the number of pixels, which keeps a transformer affordable at the full resolution of MRI images. Because the
block is shared, its size does not grow with the number of iterations: every iteration has the capacity of a large block, while the
whole model stays the size of one block.

\subsection{Latents, Adjacent Slices and Sensitivity Maps}\label{sec:inputs}
The data-consistency inputs of line~7 follow the cascade inputs of PromptMR+~\cite{xin2024promptmrplus}. PromptMR+ also passes a
learnable hidden state from one cascade to the next, to track the history of the iterations. LoopMR keeps $K$ latent images for the
same purpose. With one shared block, the latents are the block's only memory of earlier iterations beyond the estimate itself. They
are replaced by the block's output at every iteration.

LoopMR can also take adjacent slices as input, either spatially or temporally (e.g., in cine data). The sensitivity maps
(line~1) can come from any sensitivity map estimator (SME).

\section{Experiments and Results}
\subsection{Datasets and Implementation Details}
\subsubsection{Datasets}
We evaluate our model on three public datasets and one in-house dataset covering four anatomies: fastMRI (knee), Calgary-Campinas
(brain), CMRxRecon2023 (heart) and an in-house 3-D MPRAGE~\cite{mugler1990mprage} dataset (carotid).

The fastMRI dataset~\cite{zbontar2018fastmri} contains 1,594 multi-coil knee scans acquired with a 15-channel coil, and we use the
official validation set for training and testing. The validation set provides 199 volumes of proton-density (PD)-weighted knee scans. Each scan
contains about 36 slices with a matrix size close to $640 \times 368$. We partition the dataset into 159 volumes for training and
40 for testing, crop and pad each slice to $384 \times 384$ as in PromptMR+~\cite{xin2024promptmrplus}, and undersample with
fastMRI's random 1-D masks at $\times 4$ and $\times 8$.

The Calgary-Campinas (CC359) dataset~\cite{souza2018cc359,beauferris2022mcmri} provides raw multi-coil k-space of 3-D T1-weighted
brain scans from presumed healthy adults. We use its 67 fully sampled 12-coil volumes. Each volume has 256 slices with a matrix
size of $218 \times 170$ or $218 \times 180$. To make the results directly comparable with PromptMR+, we use exactly the same data split and
undersampling strategy as PromptMR+. Specifically, we split the volumes into 47 for training, 10 for validation and 10 for testing,
and use the 2-D Poisson-disc masks from the challenge at $\times 5$ and $\times 10$.

The CMRxRecon2023 dataset~\cite{wang2024cmrxrecon} provides multi-coil k-space of cardiac cine and T1/T2 mapping scans from 300
healthy volunteers. Since the reference images for the official validation and test sets are not public, we split the official training
set by subject into a training set (P001--P100, 386 scans) and a test set (P101--P120, 76 scans). Undersampling uses the
challenge's uniform 1-D masks with 24 fully sampled central lines at $\times 4$, $\times 8$ and $\times 10$. As in the challenge, every
test file is evaluated at all three acceleration factors.

The in-house carotid dataset provides multi-coil k-space of 3-D MPRAGE scans for carotid vessel wall imaging. It contains 45
volumes acquired with 6--27 receive coils at different centers. Each volume was acquired with a GRAPPA~\cite{griswold2002grappa} acceleration of 2, so the
GRAPPA reconstruction serves as the reference. We split the volumes into 32 training, 4 validation and 9 test volumes.
Undersampling uses uniform 1-D masks with 24 central lines at $\times 6$ and $\times 12$ relative to Nyquist ($3\times$ and
$6\times$ the clinical acquisition).

\subsubsection{Implementation Details}\label{sec:training}
\textit{Architecture:} We use $T = 12$ iterations, $K = 4$ latents and 5 adjacent slices (spatially adjacent or, on
the cardiac data, temporally adjacent). The block is SDUM's two-level Restormer~\cite{wang2025sdum} at half its width (channels
$[128, 256]$ instead of $[256, 512]$, 8.40M parameters, or 8.35M in the single-slice models of the ablation study, whose
block takes fewer input channels). We use the SME of
PromptMR+ (1.87M parameters). LoopMR has 10.27M parameters in total. In
the data-consistency step, $\alpha$ and $\eta$ are both initialized to 0.5.

\textit{Training:} Our models are trained with the SSIM loss on the center slice of the final output, using AdamW~\cite{loshchilov2019adamw} (learning rate
$3 \times 10^{-4}$, weight decay 0.01), gradient clipping at 0.01 and a batch size of 1 on one GPU. We follow PromptMR+'s schedules
on brain and heart (20 and 12 epochs, $\times 0.1$ decay in the last epoch) and approximate its schedule on knee (constant until
about 78\,\% of training, $\times 0.1$ after). On the in-house carotid data, we train for about 16 epochs and drop the learning rate
by $\times 0.1$ at 95\,\% of training. On brain and heart, the first 90,000 and 350,000 steps, respectively, form a pretraining
stage in which the number of iterations varies. Baselines follow their published recipes. On CC359, PromptMR and PromptMR+ use the
authors' released weights.

\textit{Evaluation:} Peak signal-to-noise ratio (PSNR) and the structural similarity index measure (SSIM)~\cite{wang2004ssim} are used as evaluation
metrics, following each dataset's official protocol. We also report the statistical significance of performance differences
using two-sided Wilcoxon signed-rank tests~\cite{wilcoxon1945} over volumes.

We implement the model in PyTorch~\cite{paszke2019pytorch} and train it on an HPC cluster with NVIDIA H100, H200 and L40S GPUs. All experiments took
over 3,000 GPU hours in total.\footnote{The code is available at \url{https://github.com/hym97/LoopMR}.}

\subsection{Comparison with State of the Art}
We compare LoopMR with three representative methods: E2E-VarNet~\cite{sriram2020e2e}, PromptMR~\cite{xin2024promptmr} (the
winner of CMRxRecon2023) and the state-of-the-art PromptMR+~\cite{xin2024promptmrplus}.

\begin{table*}[!t]
\caption{Quantitative results on the brain (CC359) and knee (fastMRI) datasets.}
\label{tab:main_bk}
\centering
\scriptsize
\setlength{\tabcolsep}{3pt}
\begin{tabular}{@{}lcccccccc@{}}
\toprule
 & \multicolumn{4}{c}{Brain (CC359)} & \multicolumn{4}{c}{Knee (fastMRI)} \\
\cmidrule(lr){2-5}\cmidrule(lr){6-9}
 & \multicolumn{2}{c}{$\times 5$} & \multicolumn{2}{c}{$\times 10$} & \multicolumn{2}{c}{$\times 4$} & \multicolumn{2}{c}{$\times 8$} \\
\cmidrule(lr){2-3}\cmidrule(lr){4-5}\cmidrule(lr){6-7}\cmidrule(lr){8-9}
Method & PSNR & SSIM & PSNR & SSIM & PSNR & SSIM & PSNR & SSIM \\
\midrule
E2E-VarNet~\cite{sriram2020e2e} & $35.01$ (0.42) & $93.64$ (0.59) & $31.91$ (0.47) & $90.09$ (0.87) & $39.03$ (1.83) & $93.28$ (2.80) & $35.06$ (2.33) & $89.25$ (3.93) \\
PromptMR~\cite{xin2024promptmr} & $36.96$ (0.41) & $94.94$ (0.54) & $34.29$ (0.38) & $92.98$ (0.58) & $40.13$ (2.04) & $93.95$ (2.91) & $36.84$ (2.45) & $90.94$ (4.10) \\
PromptMR+~\cite{xin2024promptmrplus} & $\underline{37.28}$ (0.41) & $\underline{95.13}$ (0.54) & $\underline{34.83}$ (0.35) & $\underline{93.46}$ (0.55) & $\underline{40.34}$ (2.08) & $\underline{94.10}$ (2.92) & $\underline{37.39}$ (2.48) & $\underline{91.38}$ (4.14) \\
LoopMR (ours)$^{*}$ & $\mathbf{37.45}$ (0.42) & $\mathbf{95.28}$ (0.52) & $\mathbf{35.12}$ (0.36) & $\mathbf{93.77}$ (0.53) & $\mathbf{40.44}$ (2.09) & $\mathbf{94.24}$ (2.85) & $\mathbf{37.54}$ (2.47) & $\mathbf{91.61}$ (4.10) \\
\bottomrule
\end{tabular}\\[2pt]
\parbox{\textwidth}{\scriptsize $^{*}$In every column, LoopMR differs from PromptMR+ at $p<0.01$ (two-sided Wilcoxon signed-rank test over volumes). LoopMR has a higher PSNR and SSIM than PromptMR+ on the following numbers of test volumes. Brain: 10/10 at $\times 5$ and $\times 10$. Knee: 25/25 at $\times 4$, 15/15 at $\times 8$.}
\end{table*}

\begin{table*}[!t]
\caption{Quantitative results on the heart (CMRxRecon2023) and carotid (in-house) datasets.}
\label{tab:main_hc}
\centering
\scriptsize
\setlength{\tabcolsep}{3pt}
\begin{tabular}{@{}lcccccccccc@{}}
\toprule
 & \multicolumn{6}{c}{Heart (CMRxRecon2023)} & \multicolumn{4}{c}{Carotid (in-house)} \\
\cmidrule(lr){2-7}\cmidrule(lr){8-11}
 & \multicolumn{2}{c}{$\times 4$} & \multicolumn{2}{c}{$\times 8$} & \multicolumn{2}{c}{$\times 10$} & \multicolumn{2}{c}{$\times 6$} & \multicolumn{2}{c}{$\times 12$} \\
\cmidrule(lr){2-3}\cmidrule(lr){4-5}\cmidrule(lr){6-7}\cmidrule(lr){8-9}\cmidrule(lr){10-11}
Method & PSNR & SSIM & PSNR & SSIM & PSNR & SSIM & PSNR & SSIM & PSNR & SSIM \\
\midrule
E2E-VarNet~\cite{sriram2020e2e} & $38.54$ (2.80) & $96.47$ (1.87) & $33.69$ (2.63) & $92.45$ (3.14) & $32.35$ (2.51) & $91.20$ (3.33) & $38.52$ (2.53) & $92.26$ (3.41) & $35.73$ (2.58) & $88.45$ (4.44) \\
PromptMR~\cite{xin2024promptmr} & $41.34$ (2.85) & $97.76$ (1.25) & $36.27$ (2.65) & $94.93$ (2.18) & $34.83$ (2.61) & $94.00$ (2.46) & $\underline{39.39}$ (2.94) & $93.01$ (3.37) & $\underline{37.43}$ (3.15) & $90.32$ (4.55) \\
PromptMR+~\cite{xin2024promptmrplus} & $\underline{41.66}$ (2.77) & $\underline{97.88}$ (1.17) & $\underline{36.82}$ (2.63) & $\underline{95.24}$ (2.12) & $\underline{35.55}$ (2.65) & $\underline{94.40}$ (2.49) & $39.32$ (2.91) & $\underline{93.08}$ (3.34) & $37.28$ (3.09) & $\underline{90.42}$ (4.49) \\
LoopMR (ours)$^{*}$ & $\mathbf{41.93}$ (2.73) & $\mathbf{98.00}$ (1.11) & $\mathbf{37.25}$ (2.50) & $\mathbf{95.59}$ (1.91) & $\mathbf{36.00}$ (2.57) & $\mathbf{94.85}$ (2.25) & $\mathbf{40.16}$ (3.34) & $\mathbf{93.48}$ (3.39) & $\mathbf{38.04}$ (3.46) & $\mathbf{90.91}$ (4.62) \\
\bottomrule
\end{tabular}\\[2pt]
\parbox{\textwidth}{\scriptsize $^{*}$In every column, LoopMR differs from PromptMR+ at $p<0.01$ (two-sided Wilcoxon signed-rank test over patients for the heart, subjects for the carotid). LoopMR has a higher PSNR and SSIM than PromptMR+ on the following numbers of test patients or subjects. Heart: 19/20 (SSIM) and 18/20 (PSNR) at $\times 4$, 20/20 at $\times 8$ and $\times 10$. Carotid: 9/9 at $\times 6$ and $\times 12$.}
\end{table*}

\begin{table}[!t]
\caption{Model size and inference cost over all four datasets, evaluated on one NVIDIA H100.}
\label{tab:cost}
\centering
\footnotesize
\begin{tabular}{@{}lccc@{}}
\toprule
Method & Params (M) & Latency (ms) & Peak memory (GB) \\
\midrule
E2E-VarNet & 20.1--29.9 & 29--44 & 0.3--0.8 \\
PromptMR & 79.6--81.5 & 107--158 & 2.5--9.6 \\
PromptMR+ & 92.9 & 100--121 & 1.0--3.3 \\
LoopMR (ours) & 10.3 & 203--624 & 0.7--3.0 \\
\bottomrule
\end{tabular}
\end{table}

\subsubsection{Qualitative Evaluation}
\begin{figure*}[!t]
\centering
\includegraphics[width=\textwidth]{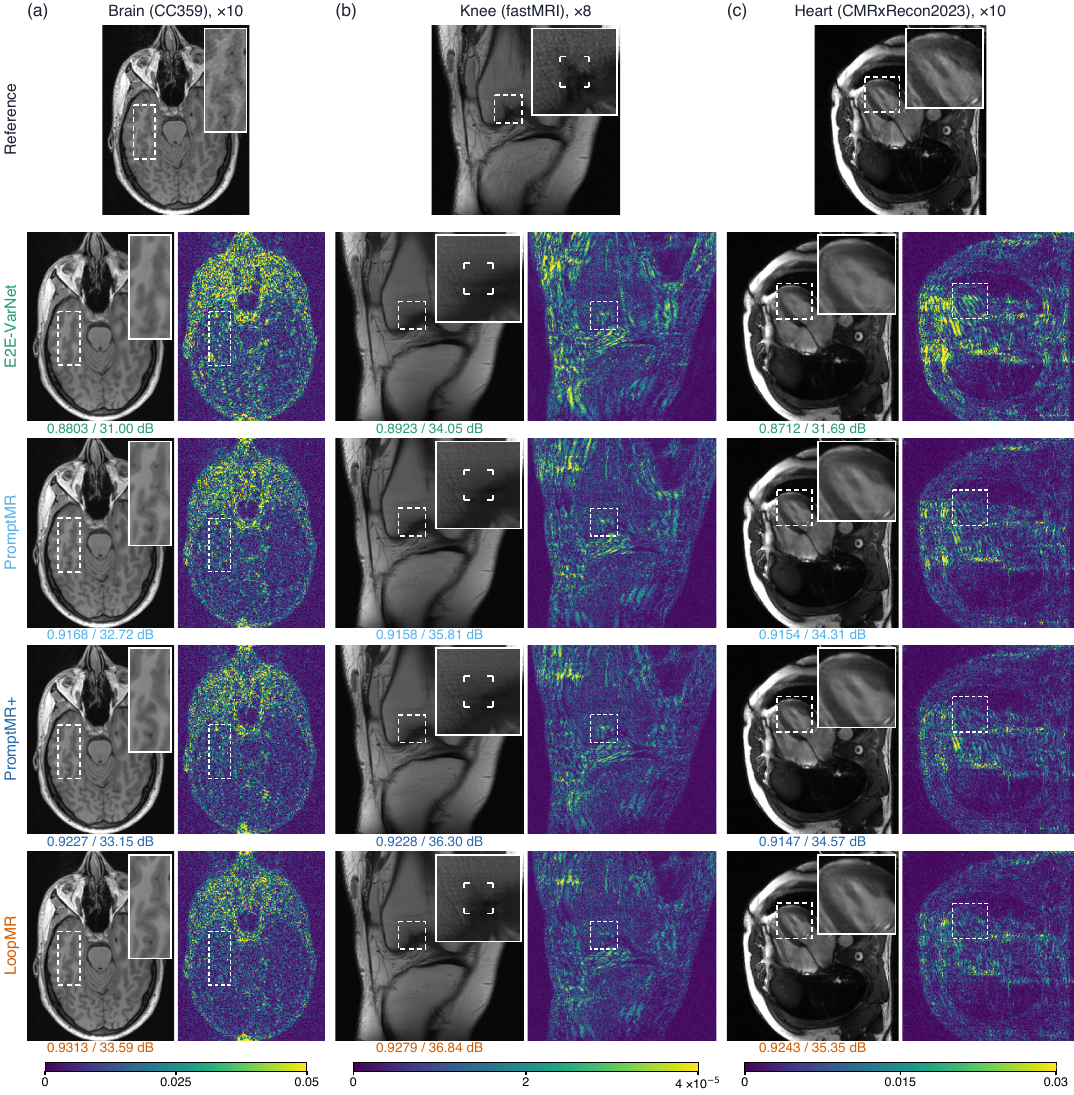}
\caption{Qualitative comparison on (a) brain (CC359, $\times$10), (b) knee (fastMRI, $\times$8) and (c) heart (CMRxRecon2023, $\times$10).
Rows: reference, E2E-VarNet, PromptMR, PromptMR+ and LoopMR. For each dataset, the reconstruction (dashed box = ROI,
enlarged in the inset) and its absolute error map are shown. Error maps: $|\hat{x} - x|$ in each dataset's own intensity units (colorbar below each dataset).
Numbers below each image: per-slice SSIM / PSNR under each dataset's reporting protocol.}
\label{fig:qualA}
\end{figure*}

\begin{figure}[!t]
\centering
\includegraphics[width=\columnwidth]{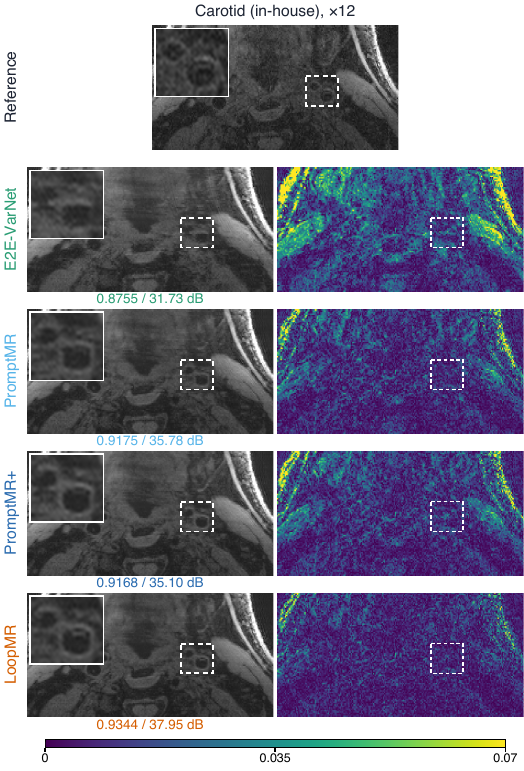}
\caption{Qualitative comparison on the in-house carotid dataset (3-D MPRAGE, $\times$12, axial slice). Layout as in Fig.~\ref{fig:qualA},
with error maps in the volume-max-normalized units of the MPRAGE evaluation.}
\label{fig:qualB}
\end{figure}

\begin{figure*}[!t]
\centering
\includegraphics[width=\textwidth]{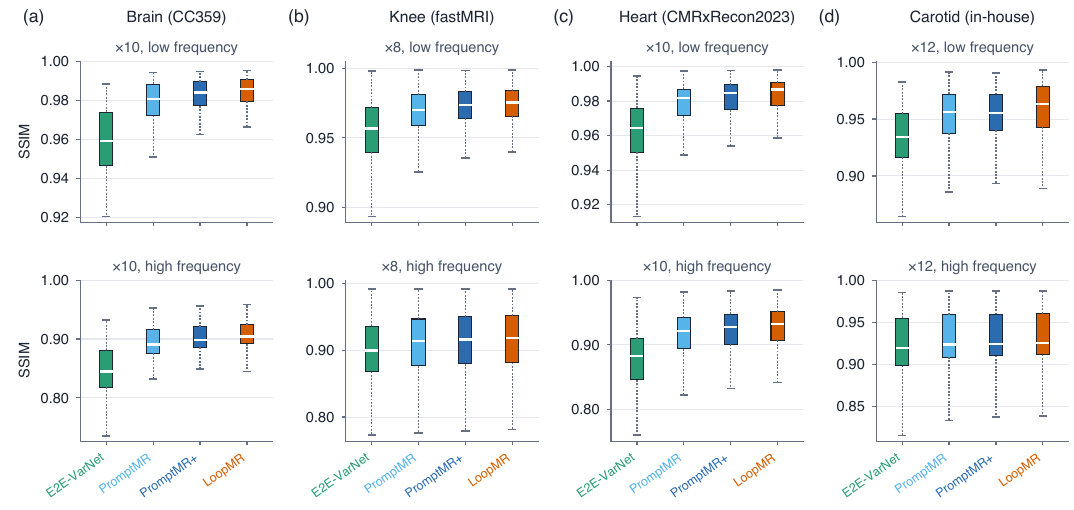}
\caption{SSIM of the low-frequency (top row) and high-frequency (bottom row) components of the reconstructions by E2E-VarNet,
PromptMR, PromptMR+ and LoopMR at the highest acceleration on (a) CC359 brain ($\times$10), (b) fastMRI knee ($\times$8), (c)
CMRxRecon2023 heart ($\times$10) and (d) in-house carotid ($\times$12). As in~\cite{guo2024reconformer}, the low-frequency component
keeps the central third of k-space along each axis and the high-frequency component the rest. Boxes summarize the per-slice SSIM
(median and quartiles), and whiskers extend to 1.5$\times$IQR. Outliers are not shown.}
\label{fig:band}
\end{figure*}

Fig.~\ref{fig:qualA} shows one case per dataset at the highest acceleration for CC359, fastMRI and CMRxRecon2023. On the
selected CC359 case (Fig.~\ref{fig:qualA}a), LoopMR reconstructs anatomical structures that are closer to the reference. In
the highlighted area, PromptMR+ even renders a temporal sulcus discontinuous, although it is continuous in the reference. On the
fastMRI dataset, we show a case with a radiologist's annotation from fastMRI+~\cite{zhao2022fastmriplus} (Fig.~\ref{fig:qualA}b).
Among all methods, LoopMR reconstructs the subchondral bone edema in this slice most faithfully. On CMRxRecon2023, LoopMR likewise
shows improved structural similarity to the reference image. Fig.~\ref{fig:qualB} shows one case from our in-house dataset.
LoopMR depicts the vessel wall closer to the reference than all competing methods.
Fig.~\ref{fig:band} shows the SSIM of the low- and high-frequency components of the reconstructions separately. The superior
performance on four different sequences across four different anatomies suggests the potential of loop transformers in
iterative refinement.

\subsubsection{Quantitative Evaluation}
Table~\ref{tab:main_bk} shows the quantitative results on CC359 ($R = 5, 10$) and fastMRI ($R = 4, 8$). Table~\ref{tab:main_hc}
shows the results on CMRxRecon2023 ($R = 4, 8, 10$) and our in-house carotid dataset ($R = 6, 12$). LoopMR outperforms all
competing methods in SSIM and PSNR on all four datasets at every acceleration factor. This holds beyond the average: LoopMR has a higher SSIM
than PromptMR+ on 137 of the 138 test cases (one volume, patient or subject at one acceleration factor), and the only exception is
one heart patient at $\times$4. Against PromptMR+, the strongest baseline, the difference is significant for
every dataset and acceleration factor ($p < 0.01$, two-sided Wilcoxon signed-rank test). The SSIM advantage of LoopMR grows with the
acceleration factor. Its SSIM gain over PromptMR+ increases from 0.15\,\% to 0.31\,\% on CC359 ($\times$5 to $\times$10), from
0.14\,\% to 0.23\,\% on fastMRI ($\times$4 to $\times$8), from 0.12\,\% to 0.45\,\% on CMRxRecon2023 ($\times$4 to $\times$10), and
from 0.40\,\% to 0.49\,\% on the carotid data ($\times$6 to $\times$12). The margin is largest on the in-house carotid data, where
LoopMR improves PSNR by 0.84 and 0.76~dB over PromptMR+ at $\times$6 and $\times$12, respectively. Table~\ref{tab:cost} lists
the model size and inference cost. LoopMR has 10.3M parameters, about 9$\times$ fewer than PromptMR+ (92.9M), at an inference time
of 203 to 624~ms per slice against 100 to 121~ms.

\subsection{Ablation Study}
\subsubsection{Shared Versus Distinct Weights}
A natural question is whether sharing weights costs performance: would distinct weights at every iteration reconstruct better?
We answer this by comparing each shared model with a distinct counterpart that differs only in its weights, under two settings:
(1) different numbers of iterations, and (2) different block families and sizes. To remove the slice-fusion effect as a
confounder, all experiments in this subsection use a single-slice model and are conducted on CC359.

\textit{Distinct-weight control:} The distinct model changes only line~7 of
Algorithm~\ref{alg:loopmr}: it uses $T$ blocks of the same architecture with separate weights, $D_{\theta_t}$ at iteration $t$,
as in a standard unrolled network. Everything else is unchanged: the inputs, the latents, $\alpha$ and $\eta$, the sensitivity
estimator and the loss.

\begin{figure*}[!t]
\centering
\includegraphics[width=\textwidth]{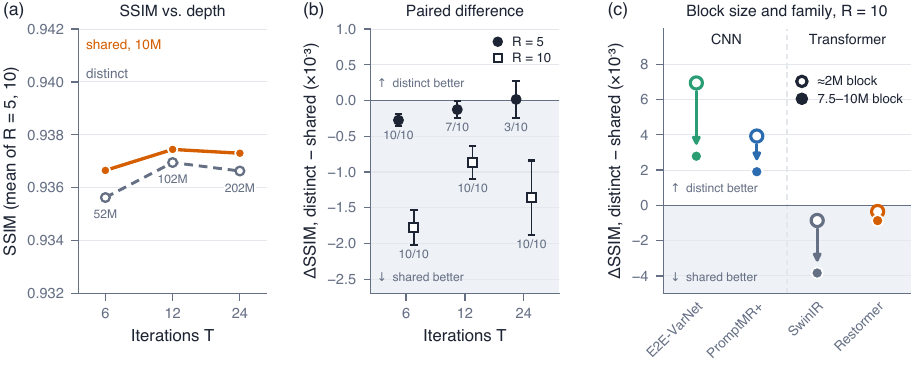}
\caption{Sharing weights across iterations, CC359 test set (10 volumes). (a) SSIM (mean of $\times$5 and $\times$10) against the number of iterations $T$
for the shared model (10.2M in total, one 8.4M block) and the distinct models (52M, 102M, 202M in total), as means over volumes. (b) Distinct minus
shared SSIM at $\times$5 and $\times$10 (mean and paired 95\,\% CI over volumes). Numbers: the number of volumes (out of 10) on which the shared model
is better. (c) Mean distinct minus shared SSIM at $\times$10 for four blocks (two CNNs and two transformers),
from a $\approx$2M block (open) to a 7.5--10M block (filled). Every pair in (c) is trained identically on both sides (12 iterations, full back-propagation,
final-output loss, gradient clip 0.01) and evaluated at 12 iterations, and the 8.35M Restormer point is the
matched $T = 12$ pair of (a) and (b).}
\label{fig:sharing}
\end{figure*}

We trained our block with shared and with distinct weights at three depths ($T = 6$, 12 and 24), as shown in
Table~\ref{tab:sharing} and Fig.~\ref{fig:sharing}(a) and (b). The SSIM (mean of $\times$5 and $\times$10) of both models grows from
$T = 6$ to $T = 12$ (from 0.9367 to 0.9375 for the shared model and from 0.9356 to 0.9370 for the distinct model) and plateaus
with more iterations (0.9373 and 0.9366 at $T = 24$). The shared model matches or outperforms its distinct counterpart at all
three depths, with 5 to 20 times fewer parameters (10.2M versus 52M, 102M and 202M). Fig.~\ref{fig:sharing}(b) shows that the
advantage of the shared model is larger at the higher acceleration: at $\times$10, it has the higher SSIM on all 10 test volumes
at every depth.

We further evaluated how the block family and the block size affect the gap. We compared shared and distinct weights for two
CNN blocks and two transformer blocks, each at two sizes (about 2M and 7.5--10M parameters per block). For the CNN family, we
evaluated E2E-VarNet (2.46M and 9.82M) and PromptMR+ (2.09M and 7.52M). For the transformer family, we evaluated
SwinIR~\cite{liang2021swinir} (2.42M and 9.56M) and our Restormer block (2.15M and 8.35M). Table~\ref{tab:sharing} and
Fig.~\ref{fig:sharing}(c) show that, at both block sizes, the distinct models in the CNN family outperform their shared
counterparts (by up to 0.0069 SSIM at $\times$10), whereas in the transformer family the shared models are better. Within each
family, a larger block shifts the balance toward sharing. The CNN gap at least halves (from 0.0069 to 0.0028 for E2E-VarNet and from
0.0039 to 0.0019 for PromptMR+), and the advantage of the shared transformer models grows.

\begin{table*}[!t]
\caption{Shared versus distinct weights on brain (CC359). S: one shared block. D: distinct blocks.}
\label{tab:sharing}
\centering
\scriptsize
\setlength{\tabcolsep}{3.5pt}
\begin{tabular}{@{}llccc cc cc cc cc c@{}}
\toprule
 & & Block & Iterations & Params (M) & \multicolumn{2}{c}{SSIM $\times 5$} & \multicolumn{2}{c}{PSNR $\times 5$} & \multicolumn{2}{c}{SSIM $\times 10$} & \multicolumn{2}{c}{PSNR $\times 10$} & S better \\
\cmidrule(lr){6-7}\cmidrule(lr){8-9}\cmidrule(lr){10-11}\cmidrule(lr){12-13}
 & Block family & (M) & S / D & S / D & S & D & S & D & S & D & S & D & ($\times 10$) \\
\midrule
\multirow{3}{*}{\rotatebox{90}{Depth}} & Restormer & 8.35 & 6 / 6 & 10.2 / 52.0 & \textbf{0.9479} & 0.9477 & \textbf{36.69} & 36.68 & \textbf{0.9254} & 0.9236 & \textbf{33.92} & 33.77 & 10/10 \\
 & Restormer & 8.35 & 12 / 12 & 10.2 / 102.1 & \textbf{0.9484} & 0.9483 & \textbf{36.76} & 36.75 & \textbf{0.9265} & 0.9257 & \textbf{34.03} & 33.98 & 10/10 \\
 & Restormer & 8.35 & 24 / 24 & 10.2 / 202.3 & 0.9481 & 0.9481 & \textbf{36.74} & 36.73 & \textbf{0.9265} & 0.9251 & \textbf{34.05} & 33.94 & 10/10 \\
\midrule
\multirow{8}{*}{\rotatebox{90}{Family}} & E2E-VarNet & 2.46 & 12 / 12 & 2.9 / 30.0 & 0.9399 & \textbf{0.9434} & 35.51 & \textbf{36.00} & 0.9073 & \textbf{0.9142} & 32.38 & \textbf{32.98} & 0/10 \\
 & E2E-VarNet & 9.82 & 12 / 12 & 10.3 / 118.3 & 0.9431 & \textbf{0.9443} & 35.93 & \textbf{36.14} & 0.9141 & \textbf{0.9169} & 32.88 & \textbf{33.21} & 0/10 \\
 & PromptMR+ block & 2.09 & 12 / 12 & 4.0 / 27.0 & 0.9444 & \textbf{0.9463} & 36.27 & \textbf{36.53} & 0.9174 & \textbf{0.9213} & 33.23 & \textbf{33.55} & 0/10 \\
 & PromptMR+ block & 7.52 & 12 / 12 & 9.4 / 92.1 & 0.9462 & \textbf{0.9472} & 36.52 & \textbf{36.68} & 0.9220 & \textbf{0.9239} & 33.62 & \textbf{33.82} & 0/10 \\
 & SwinIR & 2.42 & 12 / 12 & 4.3 / 30.9 & \textbf{0.9462} & 0.9458 & \textbf{36.44} & 36.41 & \textbf{0.9214} & 0.9205 & \textbf{33.54} & 33.49 & 9/10 \\
 & SwinIR$^\dagger$ & 9.56 & 12 / 12 & 11.4 / 116.6 & \textbf{0.9471} & 0.9461 & \textbf{36.56} & 36.46 & \textbf{0.9243} & 0.9204 & \textbf{33.82} & 33.50 & 10/10 \\
 & Restormer & 2.15 & 12 / 12 & 4.0 / 27.7 & 0.9477 & 0.9477 & 36.66 & \textbf{36.69} & \textbf{0.9249} & 0.9246 & 33.88 & 33.88 & 9/10 \\
 & Restormer & 8.35 & 12 / 12 & 10.2 / 102.1 & \textbf{0.9484} & 0.9483 & \textbf{36.76} & 36.75 & \textbf{0.9265} & 0.9257 & \textbf{34.03} & 33.98 & 10/10 \\
\bottomrule
\end{tabular}\\[2pt]
\parbox{\textwidth}{\scriptsize Depth rows: matched pairs (single slice, 4 latents, fixed iterations with full back-propagation, final-output loss, gradient clip 0.01). Family rows: every pair is trained the same way with 12 iterations and evaluated at 12; the 8.35M Restormer row is the depth pair at $T = 12$. $^\dagger$The shared model's sensitivity estimator was frozen for the first 2,000 training steps.}
\end{table*}

\subsubsection{Input Design}
\begin{table}[!t]
\caption{Ablation study of the proposed inputs and the loss of LoopMR on brain (CC359).}
\label{tab:components}
\centering
\footnotesize
\setlength{\tabcolsep}{2.4pt}
\begin{tabular}{@{}lccccc@{}}
\toprule
 & & \multicolumn{2}{c}{$\times$5} & \multicolumn{2}{c}{$\times$10} \\
\cmidrule(lr){3-4}\cmidrule(lr){5-6}
 & Params (M) & PSNR & SSIM & PSNR & SSIM \\
\midrule
Image only & 10.19 & 36.18 & 94.49 & 33.09 & 91.75 \\
+ DC inputs & 10.20 & 36.71 & 94.80 & 33.97 & 92.56 \\
+ Latents & 10.21 & 36.76 & 94.84 & 34.03 & 92.65 \\
+ 5 adjacent slices & 10.27 & \textbf{37.45} & \textbf{95.28} & \textbf{35.12} & \textbf{93.77} \\
\midrule
+ Latents, per-iterate loss & 10.21 & 36.68 & 94.79 & 33.92 & 92.54 \\
\bottomrule
\end{tabular}
\end{table}

\begin{figure}[!t]
\centering
\includegraphics[width=\columnwidth]{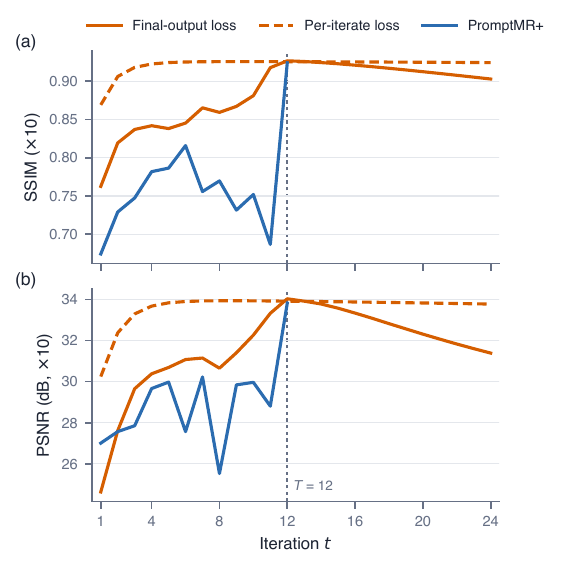}
\caption{Behavior of the loop over iterations, CC359 test set, $\times$10 (mean over the 10 test volumes). (a) SSIM and (b) PSNR
of the estimate after iteration $t$. Final-output loss: the + Latents model
of Table~\ref{tab:components}. Per-iterate loss: the same model with a loss on every iteration. Both are trained with $T = 12$
(dotted line) and run beyond it. PromptMR+: a single-slice PromptMR+ trained with the same schedule, after each of its 12 cascades (its depth is fixed at 12).}
\label{fig:loop}
\end{figure}

We evaluated the contribution of each input. Starting from the original Restormer with the image as its only input (single
slice), we added, one at a time, the three DC terms from PromptMR+, the latents and, finally, the adjacent slices. We also tested
the effect of adding a per-iterate loss to the training.

As shown in Table~\ref{tab:components}, the performance improves with each added input. Integrating the adjacent slices gives the
largest gain, 1.09~dB in PSNR and 1.12\,\% in SSIM at $\times$10. The DC terms come next, with 0.88~dB and 0.81\,\%, and the latents
add a further 0.06~dB and 0.09\,\%. Adding the per-iterate loss slightly lowers the performance at $T = 12$, by 0.11~dB and 0.11\,\%.
However, it enables anytime reconstruction, i.e., a usable image at any iteration. Fig.~\ref{fig:loop} shows the performance at
every iteration. For the model supervised only on its final output, PSNR and SSIM peak at the training depth ($T = 12$). Before
that, the intermediate iterations raise the image quality only slowly and non-monotonically: from $t = 4$ to $t = 9$, SSIM stays
between 0.84 and 0.87. After $T = 12$, both metrics decline, PSNR more sharply. By $t = 24$, the model has lost 2.64~dB in PSNR and
0.024 in SSIM. In contrast, with the per-iterate loss, both metrics improve sharply in the first few iterations. At $t = 4$, the
model is already within 0.36~dB and 0.004 of the final-loss model's peak. It also maintains its quality after $T = 12$, losing only
0.15~dB and 0.0012 by $t = 24$. This property comes at a price: its peak PSNR and SSIM (33.94~dB and 0.9255) are lower than those
of the final-loss model (34.03~dB and 0.9265).

\subsubsection{Generalizability}
\begin{table*}[!t]
\caption{Ablation study of generalizability on unseen $R$ and an unseen dataset.}
\label{tab:gen}
\centering
\footnotesize
\setlength{\tabcolsep}{4pt}
\begin{tabular}{@{}lcccccccc@{}}
\toprule
 & \multicolumn{4}{c}{Unseen $R$ (CC359)} & \multicolumn{4}{c}{Unseen data (fastMRI brain)} \\
\cmidrule(lr){2-5}\cmidrule(lr){6-9}
 & \multicolumn{2}{c}{$\times$4} & \multicolumn{2}{c}{$\times$16} & \multicolumn{2}{c}{$\times$5} & \multicolumn{2}{c}{$\times$10} \\
\cmidrule(lr){2-3}\cmidrule(lr){4-5}\cmidrule(lr){6-7}\cmidrule(lr){8-9}
Method & PSNR & SSIM & PSNR & SSIM & PSNR & SSIM & PSNR & SSIM \\
\midrule
E2E-VarNet~\cite{sriram2020e2e} & 35.31 & 93.96 & 29.37 & 85.91 & 31.33 & 84.27 & 29.00 & 79.15 \\
PromptMR~\cite{xin2024promptmr} & 37.41 & 95.22 & 31.23 & 89.54 & 36.99 & 92.45 & 33.73 & 89.18 \\
PromptMR+~\cite{xin2024promptmrplus} & \underline{37.73} & \underline{95.40} & \underline{32.10} & \underline{90.45} & \underline{37.30} & \underline{92.48} & \underline{34.57} & \underline{89.33} \\
LoopMR (ours) & \textbf{37.91} & \textbf{95.58} & \textbf{33.02} & \textbf{91.75} & \textbf{38.15} & \textbf{93.09} & \textbf{35.40} & \textbf{90.14} \\
\bottomrule
\end{tabular}
\end{table*}

To evaluate robustness~\cite{knoll2019generalization,darestani2021robustness}, we tested the CC359-trained models, without retraining, at acceleration factors not seen
in training ($R = 4$ and 16) and on an out-of-distribution multi-coil dataset, fastMRI brain~\cite{zbontar2018fastmri} (Table~\ref{tab:gen}). At the unseen
acceleration factors, LoopMR keeps its lead over the competing methods, and PromptMR+ comes second. The ranking (LoopMR, PromptMR+,
PromptMR, E2E-VarNet) is the same as at the trained accelerations. The margin over PromptMR+ grows with the acceleration factor, from
0.18~dB in PSNR and 0.18\,\% in SSIM at $\times$4 to 0.92~dB and 1.30\,\% at $\times$16. LoopMR has a higher PSNR and SSIM than PromptMR+ on all
10 test volumes at both factors. On the out-of-distribution dataset, LoopMR again outperforms all competing methods. It improves on
PromptMR+ by 0.85~dB in PSNR and 0.61\,\% in SSIM at $\times$5, and by 0.83~dB and 0.81\,\% at $\times$10 ($p < 0.001$, two-sided
Wilcoxon signed-rank test over the 186 test volumes). These margins are larger than those on the CC359 test set at the trained
acceleration factors (0.17~dB at $\times$5 and 0.29~dB at $\times$10).

\section{Discussion}
Our results revisit a long-standing design choice in unrolled MRI reconstruction. Despite some early shared-weight efforts, the field has mainly improved unrolled networks by giving each cascade its own weights, and recent challenge-winning models such as PromptMR+~\cite{xin2024promptmrplus} scale further along this path. We find that this is not the only solution. A single Restormer block of 8.4M parameters, looped 12 times, significantly outperforms PromptMR+ on all four datasets with about one ninth of its parameters (10.3M versus 92.9M). The SSIM margin grows with the acceleration factor on every dataset, so the advantage is largest where the task is hardest. In the controlled comparison, the shared model also matches or outperforms its own distinct-weight counterpart at every depth we tested. The earlier verdict against weight sharing rested on small blocks: MoDL~\cite{aggarwal2019modl} has about 0.1M parameters and ReconFormer~\cite{guo2024reconformer} 1.1M in total. Our results suggest that depth can come from iteration as well as from adding new cascades.

The ablation shows when sharing works. For the CNN blocks, distinct weights still outperform shared ones, but the gap narrows as the block grows (from 0.0069 to 0.0028 SSIM for E2E-VarNet and from 0.0039 to 0.0019 for PromptMR+), which may explain why MoDL, with its small CNN block, did not reach competitive performance. For the transformer blocks, sharing is better, and its advantage grows with the block size. We read this as evidence that distinct per-iteration weights mainly compensate for a block of limited capacity. Once one block is expressive enough to serve every iteration, separate weights add parameters without adding accuracy. This view is consistent with classical iterative reconstruction, in which a single regularizer is applied at every step. Our block is comparable in size to those of competing methods: its 8.4M parameters are similar to one PromptMR+ cascade (7.5M). Why sharing benefits the transformer blocks but not the CNN blocks remains an open question.

The input design matters as much as the block. Without the data-consistency inputs, the image-only loop oscillates between iterations and is the weakest model in Table~\ref{tab:components}. Adding the three data-consistency terms of PromptMR+ removes the oscillation and gives the largest single-slice gain. The latents add a small further gain, and the adjacent slices add the most of all. In contrast, more iterations are not a free lever. Going from 6 to 12 iterations helps, but going from 12 to 24 iterations brings limited improvement for both settings (Fig.~\ref{fig:sharing}). For the Restormer block on CC359, more than 12 iterations bring little further gain, whereas, at a fixed number of iterations, a larger block improves the accuracy (at $T = 12$, the SSIM at $\times$10 rises from 0.9249 to 0.9265 as the block grows from 2.15M to 8.35M parameters, Table~\ref{tab:sharing}). This suggests that, once a looped model reaches a moderate depth, further gains are more likely to come from a stronger block than from more iterations.

The choice of loss changes how the loop behaves (Fig.~\ref{fig:loop}). With the loss on the final output, which we use for LoopMR, the reconstruction is best at the trained depth, but the intermediate estimates are of low quality and the quality falls when the loop runs past $T = 12$. At $\times$10, SSIM stays between 0.84 and 0.87 for most of the loop, and by $t = 24$ the model has lost 2.64~dB in PSNR. A loss on every iteration gives a model that is close to its final quality after four iterations and stays stable up to 24, at a cost of about 0.1~dB at the trained depth. PromptMR+ behaves like our final-loss model, and its depth is fixed at 12 by construction. A shared block, in contrast, can be run for any number of iterations. This opens the way to choosing the depth at inference, for example a short loop for a fast preview. Such a preview could be clinically useful: the operator could check coverage, positioning and motion while the scan is still running, and decide on a repeat before the patient leaves, while the same network completes the full-quality reconstruction afterwards.

LoopMR also generalizes well. At accelerations not seen in training, its margin over PromptMR+ grows with the acceleration factor, from 0.18~dB at $\times$4 to 0.92~dB at $\times$16. On fastMRI brain, the margins are larger than on CC359 itself.

Our study has limitations. First, the saving is in parameters, not in time: LoopMR runs its compute-heavy block 12 times, and its inference takes 203 to 624~ms per slice against 100 to 121~ms for PromptMR+ (Table~\ref{tab:cost}). Since accuracy depends on the block rather than on the number of iterations, developing a more efficient block that keeps this accuracy at a lower cost is an important direction for future work. Second, the training sets are moderate (47 training volumes on CC359), and whether distinct weights regain an advantage with much larger multi-site training data remains unknown. Third, each model was trained for one dataset and with a single family of sampling masks (2-D Poisson-disc masks on CC359 and 1-D Cartesian masks on the other datasets), so we do not expect it to generalize to unseen anatomies or sampling patterns. Recent cardiac challenges such as CMRxRecon2024~\cite{wang2025cmrxrecon2024} gain this robustness by training on many mask types, and LoopMR could adopt the same strategy. Finally, all undersampling is retrospective, and the in-house carotid data are not public. Image quality was also assessed with PSNR and SSIM rather than by radiologists, so possible hallucinations at high acceleration remain to be evaluated. Future work will focus on making LoopMR deployable in the clinic: reducing its latency while keeping its memory footprint small, and evaluating it prospectively with a reader study.

\section{Conclusion}
In this study, we revisited weight sharing in unrolled MRI reconstruction. Inspired by looped transformers in large language
models, LoopMR applies one shared Restormer block at every iteration, together with data-consistency inputs, latents and adjacent
slices. It significantly outperformed the challenge-winning PromptMR+ on four datasets (brain, knee, heart and carotid), with about one ninth of its parameters. In a controlled comparison, the shared model matched or outperformed
its distinct-weight counterpart at every depth. Sharing also favored the transformer blocks increasingly as the block grew,
whereas the CNN blocks still benefited from distinct weights. These results show that depth in unrolled reconstruction can come
from iteration rather than only from new weights. They suggest that looped transformers are a promising direction for MRI
reconstruction and, with further work on efficiency and clinical validation, for clinical use.

\providecommand{\refname}{References}

\end{document}